\documentclass[a4paper]{jpconf}
\usepackage{graphicx}
\begin{document}
\title{Scalar Mesons and FAIR}

\author{Denis Parganlija}

\address{Institute for Theoretical Physics, Vienna University of Technology, Wiedner Hauptstr.\ 8-10, A-1040 Vienna, Austria}

\ead{denisp@hep.itp.tuwien.ac.at}

\begin{abstract}
I review issues related to the scalar-meson phenomenology considered of interest for the future PANDA experiments at FAIR.
\end{abstract}

\section{Introduction}

Scalar mesons are degrees of freedom of QCD (Quantum Chromodynamics) that possess quantum numbers $J^{PC} = 0^{++}$,
where $J$ is the total spin, $P$ represents parity and $C$ is the charge 
conjugation. Mesons that possess isospin $I=0$ are referred to as isoscalars. Scalars with $I=0$ represent QCD degrees of freedom with some 
peculiar features: they are defined as carrying quantum numbers
of vacuum but are 
identifiable as resonances; as mesons, they should be of $\bar q q$ (quarkonium) structure but are too many to simply 
represent $\bar q q$ objects; and they are strongly mixed with each other so that disentangling them is in the majority of 
cases a highly non-trival task.
However, they are also of extreme importance for QCD since their vacuum expectation values represent the
cause of the spontaneous breaking of the chiral symmetry in QCD, and this "Higgs mechanism" is in turn expected to generate 
Goldstone bosons of QCD: pions, kaons and others. Thus understanding isoscalar mesons
is a fundamental question of QCD, and one that has by no means been closed after decades of research and debate.
\subsection{Scalar Mesons and Experiment}

Listings of the Particle Data Group (PDG) cite the existence of 
five $IJ^{PC} = 00^{++}$ states in the energy region up to 1.8 GeV: $f_0(500)$ or $\sigma$, $f_0(980)$, $f_0(1370)$, $f_0(1500)$ and $f_0(1710)$ \cite{PDG}. Let us review 
basic experimental data regarding these resonances.
\begin{itemize}

\item Historically, debates regarding scalar mesons have been mostly focused on the $f_0(500)$ resonance (also known as 
$\sigma$). The latter
is characterised by a decay width that is of the same magnitude as the resonance mass -- according to PDG, it has a mass of (400-500) MeV and 
a decay width of (400-700) MeV. The 
existence of this state was suggested in linear sigma models long ago: the state was introduced
theoretically as the putative chiral partner of the pion \cite{gellmanlevy}. Since the pion is a 
well-established quarkonium state, then the chiral-partner state of the pion had to be of the same structure. Additionally, the pion is a
pseudo-Goldstone bosone -- its mass is the lowest in the meson spectrum. 
Therefore, naive expectations were that ({\it i}) the chiral partner of the pion would be the lightest scalar state with $I=0$, 
i.e., the $\sigma$ meson, and consequently ({\it ii}) that the $\sigma$ meson was a $\bar q q$ state. However, currently it is by no means clear that 
$\sigma$ is truly a $\bar q q$ state, implying that its interpretation as the chiral partner of the pion is in doubt.
We will come back to this issue later in this manuscript.

\item First hints about the existence of the $f_0(980)$ resonance were discovered in analyses of ${\pi\pi}$ scattering data
found to exhibit a rapid drop in the cross-section in the energy region
between $950$ MeV and $980$ MeV (i.e., close to the $KK$ threshold) in Saclay data on antiproton-proton 
\cite{Bizzarri:1970} and Berkley data on pion-proton collisions \cite{Berkley}. The most precise modern-data analyses result 
in pole mass and pole decay widths values in the close vicinity of the mentioned energy region -- see, e.g., Ref.\ \cite{Kaminski}
where the pole mass was determined as $m_{f_0(980)} = (996 \pm 7)$ MeV and the pole decay width as $\Gamma_{f_0(980)} =
50^{+20}_{-12}$ MeV.

\item The $f_0(1370)$ resonance is a broad enhancemenet in the $\pi\pi$, $4\pi$ and, to a lesser extent, $KK$ and $\eta \eta$ 
channels in the energy region of approximately 1.3 GeV. It is characterised by emergence of two peaks, respectively in the 
$2\pi$ and $4\pi$ decay channels \cite{buggf0}, the disentaglement of which requires a simultaneous analysis of $\pi\pi 
\rightarrow \pi\pi$ and $\pi \pi \rightarrow 4\pi$ scattering data. The analysis of Ref.\ \cite{buggf0} demonstrates that the 
$2\pi$ 
decay channel is dominant in the energy region up to 1.35 GeV whereas the energy region thereafter is dominated
by the opening of the $\rho \rho$ threshold leading to a strong $4\pi$ decay pattern.
Note that $f_0(1370)$ represents a single resonance despite possessing two peaks since a combined dispersive 
analysis of both $2\pi$ and $4\pi$ channels yields only one pole. 
Historically, early hints of the $f_0(1370)$ existence were observed in $\bar{p} p$ collisions at CERN in 1969 with a 
possible $\rho\rho$ enhancement claimed at $1.4$ GeV \cite{Donald:1969}. Current edition of the Particle Data Group listings
accumulates all available data on $f_{0}(1370)$ estimating an interval for, rather than stating exact values of, its mass and 
decay width: $m_{f_{0}(1370)}=(1200$-$1500)$ MeV and $\Gamma_{f_{0}(1370)}=(200$-$500)$
MeV \cite{PDG}.

\item The discovery of the $f_{0}(1500)$ resonance originated in search for the
scalar glueball state. This resonance is found mostly in pion final states
from nucleon-nucleon (or antinucleon-nucleon) and pion-nucleon scattering
processes. If such processes produce four pions, then $f_{0}(1500)$ is
reconstructed from $\rho\rho$ final states in the $2(\pi^{+}\pi^{-})$ channel
and from $\sigma\sigma$ final states in the $2(\pi^{+}\pi^{-})$ or $2(\pi
^{0}\pi^{0})$ channels. The resonance is therefore at least partly
reconstructed in channels containing a double Pomeron exchange rendering the
state a glueball candidate \cite{Close:1987}.
The PDG cites a world-average mass $m_{f_{0}(1500)}=(1505\pm6)$ MeV and decay
width $\Gamma_{f_{0}(1710)}=(109\pm7)$ MeV \cite{PDG}.

\item The $f_0(1710)$ resonance is characterised by a predominant $KK$ decay channel, marking a clear point of 
distinction between this resonance and the above mentioned ones.
The earliest evidence for the $f_{0}(1710)$ resonance was obtained from the
decay $J/\psi\rightarrow\gamma\eta\eta$ at the SLAC Crystal Ball detector from
$e^{+}e^{-}$ annihilation \cite{f0(1710)-1981-SLAC}; a
resonance with a mass of $(1640\pm50)$ MeV, a decay width of $220_{-70}^{+100}$ MeV
and the charge conjugation
$C=+1$ was found (the latter was because the resonance had been produced in a radiative $J/\psi$ decay but, interestingly, no 
final determination of the total spin and parity was possible from these first data).
Subsequent analyses, such as those in Ref.\ \cite{f0(1710)}, determined the resonance to be of $J^{P}=0^{+}$ nature.
The PDG cites a world-average mass $m_{f_{0}(1710)}=(1720\pm6)$ MeV and a decay
width $\Gamma_{f_{0}(1710)}=(135\pm8)$ MeV \cite{PDG}.
\end{itemize}

\subsection{Scalar Mesons and Theory}

Assuming a $\bar{q} q$ structure for mesons (where $q$ denotes a constituent quark) it is possible to construct two $IJ^{PC} 
= 00^{++}$ states for three quark flavours
(up $u$, down $d$ and strange $s$) once the approximate isospin symmetry in the nonstrange-quark sector is considered ($u = d$):
$
f_0^N \equiv (\bar{u}u+\bar{d}d)/\sqrt{2}$ and $f_0^S \equiv \bar{s}s$. The latter are per construction pure states. They can obviously mix by virtue of carrying the same quantum numbers; the 
mixed states are expected to be those
appearing in the physical spectrum. Clearly, two pure states will produce the same number of mixed states -- thus only two of the five previously discussed physical states
can conceivably be described in this approach. Hence the natural questions are

\begin{itemize}

\item Where in the physical spectrum are isoscalar $\bar{q} q$ states located?

\item What structure do the remaining states possess? 

\end{itemize}

Answers to these questions are complicated by the already mentioned fact that states with the same $IJ^{PC}$ quantum
numbers in general, and isoscalar mesons in particular, are expected to mix. Thus an isoscalar (or indeed any other) meson need not necessarily be of
$\bar q q$ structure -- it can also possess admixtures (or even be predominantly composed) of tetraquark ($\bar{q}\bar{q}qq$),
glueball (bound states of QCD gauge bosons -- gluons) or molecular (e.g., $\pi\pi$ or $KK$) contributions. For the cases of the above-stated isoscalar mesons we then note the following:

\begin{itemize}
\item As indicated above, the $f_0(500)$ resonance is usually considered as the natural option for the chiral partner of the pion, implying that $f_0(500)$
possesses the $\bar{q} q$ structure
as suggested, e.g., by Ref.\ \cite{NJL}. However, a $\bar
{q}q$ scalar state possesses the intrinsic angular momentum $L=1$ as well as
the relative spin of the quarks $S=1$. For this reason one could also easily
expect the scalar $\bar q q$ state to be in the region above $1$ GeV.
This would imply that ({\it i}) isoscalars above 1 GeV possess the $\bar q q$ structure and ({\it ii}) $f_0(500)$ represents a tetraquark state or a $\pi\pi$ 
bound state. For details on both statements see, e.g., Refs.\ \cite{Meinereferenzen, Meinereferenzen1, Paper3}.

\item As already discussed, the $f_0(980)$ resonance is close to the kaon-kaon threshold rendering a structure analysis for this state rather difficult. The resonance may be interpreted as a quarkonium
\cite{f0(980)asqq}, as a $\bar{q}^{2}q^{2}$ state
\cite{f0(980)asq2q2}%
, as a $KK$ bound state
\cite{f0(980)asKK}%
, as a glueball \cite{f0(980)-as-glueball} or even as an $\eta\eta$ bound
state \cite{f0(980)etaeta-f0(1500)glueball-f0(1370)-f0(1710)qq}.

\item The $f_0(1370)$ resonance appears to be a suitable candidate for a non-strange $\bar{q} q$ state \cite{Meinereferenzen1,Paper3} or a dynamically generated state
\cite{Oset}.

\item The $f_0(1500)$ resonance seems to represent a (predominant) glueball state \cite{Close:1987, Stani} although
claims have also been made that, contrarily, $f_0(1710)$ is 
of predominantly glueball nature \cite{f0(1710)asglueball}. Alternatively, $f_0(1710)$ can also be interpreted as a 
predominantly $\bar s s$ state
\cite{Paper3}.

\end{itemize}

Hence the current state of knowledge allows us to only claim with certitude that all the above mentioned states mix \cite{Schechter}; some of them even overlap due to large decay widths \cite{PDG}.
The situation is thus rather complicated -- and it is enhanced even further by the possible existence of a sixth isoscalar state, located very proximal to the set
of the already mentioned isoscalars: the $f_0(1790)$ resonance.

\section{The \boldmath $f_0(1790)$ Resonance and FAIR}

The existence of the putative new $f_0(1790)$ resonance has been claimed by the BES II Collaboration in 2004 \cite{BESII2004}. The mass and decay width of the 
resonance were determined by the Collaboration as $m_{f_{0}(1790)}=1790_{-30}^{+40}$ and $\Gamma_{f_{0}(1790)}=270_{-30}^{+60}$ MeV. The stated large decay width implies a strong overlap
of this resonance with $f_0(1710)$; however, there is a clear point of distinction between the two resonances: $f_0(1710)$ is reconstructed predominantly in the kaon
decay channels whereas $f_0(1790)$ is reconstructed predominantly in the pion decay channels.\\
There are four basic production mechanisms for $f_0(1710)$ and $f_0(1790)$ via $J/\psi$ decays \cite{BESII2004}:

\begin{itemize}

\item ({\it i}) $J/\psi\rightarrow\varphi K^+ K^-$, 

\item ({\it ii}) $J/\psi\rightarrow\varphi \pi^{+}\pi^{-}$,

\item ({\it iii}) $J/\psi\rightarrow\omega K^{+}K^{-}$, 

\item ({\it iv}) $J/\psi\rightarrow\omega \pi^{+}\pi^{-}$.

\end{itemize}

Reactions ({\it i}) and ({\it iii}) allow for reconstruction of $f_{0}(1710)$ \cite{PDG}
whereas reactions ({\it ii}) and ({\it iv}) allow for reconstruction of $f_{0}(1790)$.
Importantly, assuming $f_{0}(1710)$ and $f_{0}(1790)$ to be the same resonance leads to a contradiction: 
such a resonance would have to possess a pion-to-kaon-decay ratio of $1.82 \pm 0.33$ according to reactions
({\it i}) and ({\it ii}) and, simultaneously, the pion-to-kaon-decay ratio $< 0.11$ according to reactions
({\it iii}) and ({\it iv}) \cite{BESII2004}. There can be no one resonance with these simultaneous features -- thus
BES II data suggest $f_{0}(1710)$ and $f_{0}(1790)$ to represent two distinct resonances.\\

Let us finally discuss the relevance of the deliberations in this paper for FAIR. The proposed PANDA experiments at FAIR will study
interactions between antiprotons and fixed-target protons and nuclei in the momentum range of 1.5-15 GeV. A stated goal of the PANDA Collaboration is the
exploration of light-hadron spectroscopy, including search for glueballs and multiquark states \cite{PANDA}. In this regard I 
would like to suggest the following courses of action in the 
particular case of possible mixing scenarios in the scalar sector:

\begin{itemize}
\item Any viable search for members of a glueball spectrum must include the ground state as otherwise the spectrum would be incomplete.
The PANDA Collaboration is in a unique position to greatly advance our search for glueballs since ({\it i}) the Collaboration can build upon extensive experimental work
performed in the last decades and ({\it ii}) the technical specifications of the PANDA detector allow for a search for glueballs with unprecedented accuracy in
comparison with similar undertakings in the past.\\
However, the glueball state will inevitably overlap with other states that possess the same quantum numbers (and may be of $\bar q q$, $\bar{q}\bar{q}qq$ or molecular
structure) and the disentanglement of various signals from each of the mentioned resonances will represent a formidable task.\\

\item Current theoretical results \cite{Stani,f0(1710)asglueball} suggest the scalar glueball to be positioned at (1.5-1.7) GeV, i.e., in the vicinity of the 
established resonances $f_0(1370)$, $f_0(1500)$ and $f_0(1710)$ as well as the newly proposed $f_0(1790)$ state. Consequently, a viable search for the scalar glueball
will have to consider overlap of the glueball not only with $f_0(1370)$, $f_0(1500)$ and $f_0(1710)$ \emph{but also} with $f_0(1790)$. In other words:
a viable search for the glueball ground state at PANDA must consider the possibility of the existence of $f_0(1790)$.
If $f_0(1790)$ is not found at PANDA, then BES results may be put in doubt; but if $f_0(1790)$ is ascertained then
its signal will have to be disentangled from the glueball signal in order for the glueball search to be successful, 
making the search even more challenging and thus even more interesting.

\end{itemize}

\section{Conclusions}

Scalar mesons represent an interesting challenge for the future PANDA experiments at FAIR: the planned search for the 
glueball state in the
$IJ^{PC} = 00^{++}$ channel will have to consider mixing/overlap between non-strange $\bar q q$, strange $\bar q q$, non-strange and strange $\bar{q}\bar{q}qq$
as well as possible molecular-type states. Additionally, obtaining a clear signal for the glueball will inevitably require the confirmation (or negation)
of the existence of a putative new isoscalar state referred to as $f_0(1790)$ by the BES II Collaboration \cite{BESII2004}. If $f_0(1790)$ exists, it will most certainly overlap with the glueball. Thus
a viable search for a glueball state simultaneously implies a search for $f_0(1790)$: ignoring $f_0(1790)$ will strongly distort
any signal for the glueball. For this reason, the search for a glueball at PANDA is more than a search for merely one state -- it is a search where several very
close-by states have to be considered simultaneously.

\end{document}